# FPGA Based Design And Implementation of Neural Network Compensators for Nanocrystalline MEMS Pressure Sensor

Somsubhra Talapatra somsubhra@pdsit.becdu.ac.in

Sudip Ghosh sudip_etc@yahoo.co.in

C Roy Chaudhuri chirosreepram@yahoo.com

Biplab K Sikdar biplab@cs.becs.ac.in

Hafizur Rahaman rahaman_h@yahoo.co.in

Ganapati Sengupta VLSI Laboratory
Bengal Engineering & Science University, Shibpur
Howrah-3, India 711103

## Abstract

***In this paper the effects of nonlinearity and temperature of nanocrystalline MEMS pressure sensor have been compensated using neural network algorithm which has been implemented on a FPGA chip. The neural network employs back propagation algorithm to train the multilayer perceptron feed forward network that causes an error reduction due to temperature upto the level of 98%. It is realized on FPGA with pipelined architecture for parallel and fast sensing operation. The pressure sensor along with the FPGA based compensation results in an intelligent sensor system capable of high speed and accurate pressure measurement which has been tested for a typical biomedical application .***

## 1. Introduction

Nanocrystalline silicon based MEMS pressure sensor has been recently gaining importance for improving the performance of pressure sensors due to its increased piezoresistive coefficient [1]. It is found that the pressure sensitivity of such nanocrystalline silicon pressure sensor is three times better in comparison to the conventional silicon pressure sensors for a particular nanowire dimension [2,3]. However, such sensors also possess nonlinearity and considerable temperature effects, which need to be compensated for efficient utilization of a sensitive pressure sensor [4].

In this paper, the compensations have been achieved by neural network techniques using multiplayer perceptron feedforward network with signal processing neurons in its hidden and output layers. The number of neurons in the hidden layer and the weights have been optimized using back propagation algorithm for a minimum error. The training of the ANN has been done offline using MATLAB ANN toolbox. The trained network is then implemented on a FPGA chip through introduction of pipelined architecture leading to efficient utilization of the FPGA resources unlike the simple behavourial architecture. This enables parallel operations resulting in improved speed. The datapath for this pipeline is optimized for area so that the entire system can be implemented as an ASIC where minimization of silicon area is very important. The scheme has been implemented on Xilinx Spartan chip which has relatively lesser number of combinational logic blocks and this has been made possible due to the optimized pipeline architecture.

## 2. Sensor Design, Fabrication and Results

Nanocrystalline silicon MEMS based pressure sensors are being used for low pressure range biomedical applications such as espirators, ventilators,spirometers typically in the range of 0–1 bar Typical specifications for a pressure sensor used in respirators or for blood pressure measurement are as follows [5]:

- pressure range: 0–1 bar;
- sensitivity desired: 30 mV $V^{-1}$ $bar^{-1}$;
- nonlinearity: less than 0.5% of full-scale output;
- resolution: 1 mbar and
- operative temperature range: 25 $^{o}$C–80 $^{o}$C.

The design and fabrication of the nanocrystalline MEMS pressure sensor have been reported in details elsewhere [4]. The schematic of such sensor along with its pressure and temperature characteristics are shown in Fig.1 and 2 respectively. It can be observed from Fig.2. that the pressure sensor exhibits significant temperature dependence which needs to be compensated.

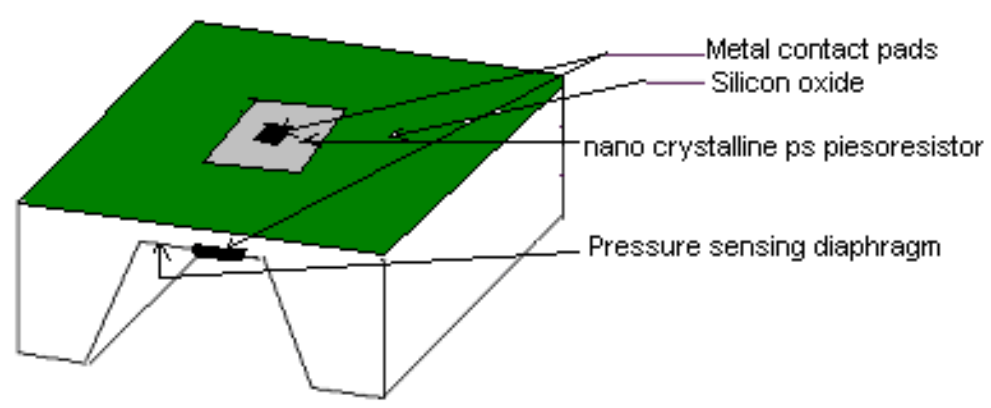

**Figure 1. Cross sectional view of the sensor**

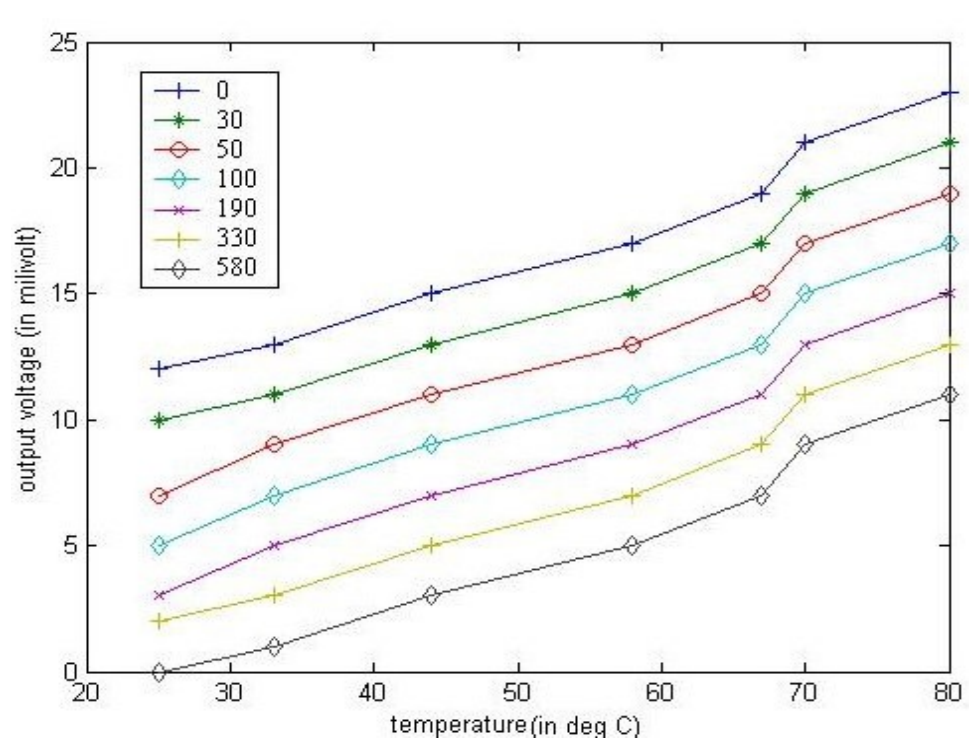

**Figure 2. Variation of output voltage of pressure sensor with temperature for different pressure values**

## 3. Compensation using ANN

The compensation of the pressure sensor has been done using ANN. The necessary set of data for input output pairs in the inverse ANN model are obtained from the experiments. The proposed scheme, based on inverse model, for compensation is shown in Fig. 3. The actual sensor and ANN are connected in cascade. The sensor output is fed as one of the inputs of the ANN and ambient temperature as the other input.The ANN is based on mutilayer perceptron (MLP) feed forward network with signal processing neurons in its hidden and output layers. It has three layers such as input, hidden and output layer. The input layer is having three inputs including bias and the output layer has one output. The number of neurons in the hidden layer is selected heuristically.

The network has been trained starting with minimum number of neurons in hidden layer for a fixed epoch. The number of neurons which give minimum error has been selected as the optimum neurons in the hidden layer, where the activation function in hidden layer is tansigmoidal and the output layer is linear. The well-known back propagation algorithm, which is a generalization of the LMS (least mean square) algorithm, is used to train the MLP [6].

### 3.1 Simulation Studies

The simulation studies are carried out in MATLAB, ANN toolbox environment.

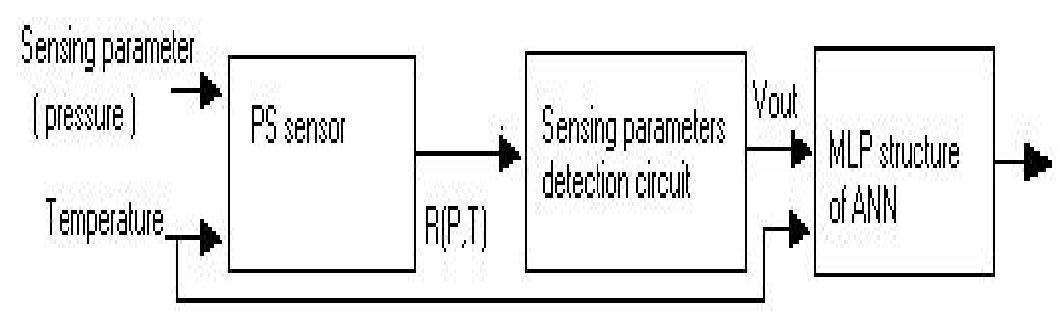

**Fig 3: Schematic of the proposed model**

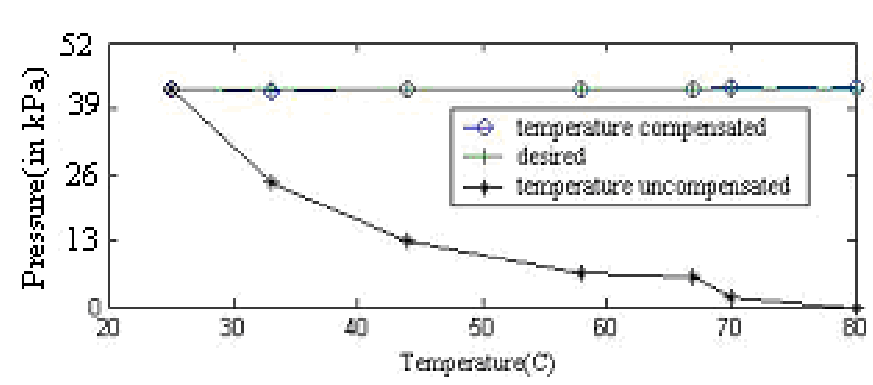

**Fig.4. Variation of pressure with temperature for the compensated and uncompensated sensor**

All the parameters of the sensor such as pressure, ambient temperature and voltage output of the detection circuit have been normalized so as to keep their values within {-1, 1}. Seven data sets, from the graph shown in Fig.2, corresponding to seven pressure values are generated for training as well as testing of the ANN model. Six data sets are chosen randomly for training and one for testing the trained network. The optimum MLP structure obtained heuristically is 3-12-1.The parameters for training the MLP are learning rate 0.1, error goal 0.00001 and total epochs 5,000. Each set of input pattern is applied individually to update the weights of the network. Completion of all patterns of all the training sets form one epoch of training. The mean squared error (MSE) in the network reduces with the progress in training. After training, the final weights are stored in the ANN model for testing and performance evaluation of the network with test patterns. The testing of the network is done to find out whether the network can estimate/predict the response of the sensor for any input within the specified temperature and pressure for which it is trained. Fig.4 shows the temperature effect of compensated pressure output at different temperature corresponding to a desired pressure of 40kPa. The effect of temperature compensation by the ANN network is clearly visible and compared with the uncompensated output. The trained weights after MATLAB simulation is shown in Table 1.

**Table 1 Final trained weights of ANN**

| W | | | $U^T$ |
|---|---|---|---|
| 3.23 | -1.64 | 2.30 | -2.38 |
| -237.78 | 105.32 | 95.37 | 0.04 |
| 87.35 | 50.21 | 122.13 | -0.18 |
| 0.49 | 31.14 | -7.39 | -696.20 |
| 0.49 | 11.74 | -5.43 | -22.92 |
| -0.43 | -11.32 | 3.40 | -725.35 |
| 1.78 | -8.67 | 0.70 | 6.23 |
| 3.28 | -1.73 | 0.56 | -0.36 |
| | | | 1.95 |

The following section includes the description of the hardware implementation of the pipelined ANN architecture along with the designed pipelined timing cycles .

## 4. FPGA Based Implementation

### 4.1 Motivation for Pipelined Architecture Design for the above ANN

This section reports the FPGA based implementation of the ANN compensator described in the earlier sections, to eliminate the effects of nonlinearity and temperature. The structure of the ANN ,that is used here , is organized in 2 layers and is shown in the Fig.5 The first layer has 12 neurons and the second layer has only one neuron. Each of the first layer neurons takes three weighted inputs, one for pressure and one for temperature data and one for a biasing input. The sum of these inputs is passed through a tansigmoidal activation function,as given in equation 2. All the pressure inputs and all the temperature inputs of all the first layer neurons are connected to the pressure and temperature inputs of the system respectively, as shown in Fig5. The 12 outputs from the first layer and a dedicated bias input are multiplied by proper weights and added in the second layer neuron. The 2nd stage neuron does not use any activation function to produce final output.

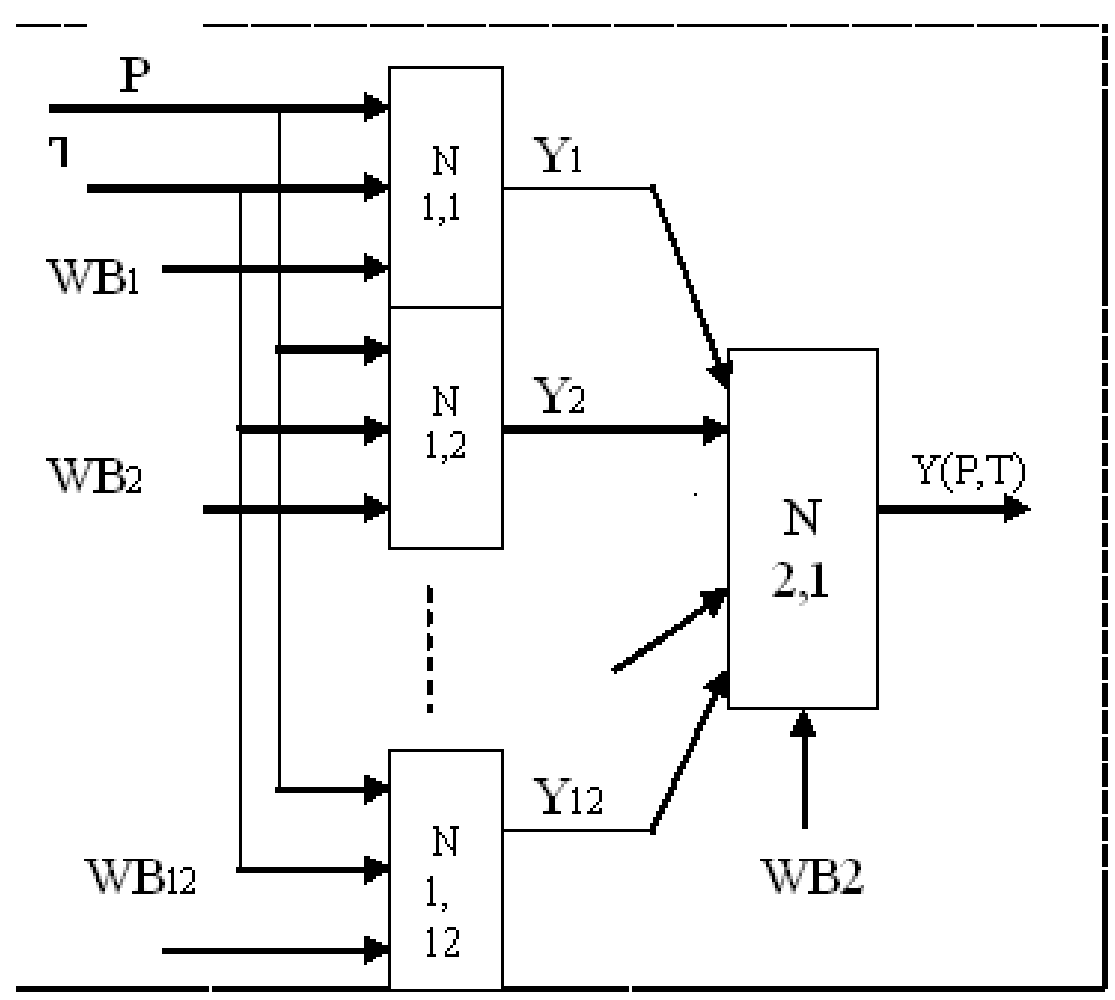

**Figure 5: 2 Layered Structure of the ANN.**

The ANN's organization can be expressed mathematically as given in equations 1and 2, where

$$Y(P,T) = WB2 + \sum_{i=1}^{12} W2_i \times Tansig(WP_i \times P + WT_i \times T + WB_i) \quad \ldots(1)$$

$$Tansig(x) = \frac{e^{x} - e^{-x}}{e^{x} + e^{-x}} \quad \ldots(2)$$

*P* is the normalised pressure input, *T* is Temperature input, *WPi* and *WTi* are weights associated with *P* and *T* respectively in *i'th* neuron in the first layer, *WBi* is the biasing weight of the *i'th* neuron in the 1st layer, *W2i* are the weights for the 2nd layer's input and *WB2* is the biasing weight of the 2nd layer. Equation 1 can be broken down in smaller iterative equations as shown in equation 3. Except for the addition of $Y_{i-1}$ the rest of the equation can be broken down into 12 independent equations 4 or segments and can be calculated in parallel. This gives rise to the possibility that equation 3 can be implemented using a pipelined architecture which will be faster than a sequential architecture. The timing diagram of the proposed pipeline for equation 3 is given in Fig6.

$$\left.\begin{aligned} &Y_0 = WB2 \\ &Y_i = Y_{i-1} + W2_i \times Tansig(WP_i \times P + WT_i \times T + WB_i), for \quad i = 1,2,\dots 12 \\ &Y(P,T) = Y_{12} \end{aligned}\right\} \quad (3)$$

$$Y_i' = W2_i \times Tansig(WP_i \times P + WT_i \times T + WB_i), for \quad i = 1,2,\dots 12 \quad ..(4)$$

**4.2 Design and implementation of the pipeline architecture**

The pipeline involves 3 multiplications and 3 additions. The datapath for this pipeline is optimized for area so that the entire system can be implemented as an ASIC where minimization of silicon area is very important.

| T1 | T2 | T3 | T4 | T5 | T6 | T7 | T8 | T9 | T 10 | T 11 | T 12 | T 13 | T 14 | T 15 |
|---|---|---|---|---|---|---|---|---|---|---|---|---|---|---|
| P1 | P2 | P3 | P4 | P5 | P6 | P7 | P8 | P9 | | | | | | |
| | | | P1 | P2 | P3 | P4 | P5 | P6 | P7 | P8 | P9 | | | |
| | | | | | | P1 | P2 | P3 | P4 | P5 | P6 | P7 | P8 | P9 |

**Figure 6: ANN PIPE Controller**

**T1-T15** represent the timing cycles

**P1**: Fetch 1: Address throw cycle to ROMs $(WP_{i,} WT_i, Z = WB_i)$

**P2**: Fetch 2: ROM access cycle $(WP_{i,} WT_i, Z = WB_i)$

**P3**: Mult 1: $X <= P \times WP_i$

**P4**: MultAdd: $Y <= T \times WT_i, AA <= X + Z$

**P5**: Add2: *A<=AA + Y*

**P6**: Tansig1: Address Throw to ROMs (Tansig LUT, $W2_i$ )

**P7**: Tansig2: ROM access cycle; F <= Tansig data, $W2 <= W2_i$

**P8**: Mult3: *G1<=F x W2*

**P9**: Add3: *G<= G + G1*

The datapath shown in Fig7 contains only one 16bit signed multiplier and one 16bit adder and as a result of this, successive P1 stages are issued with a delay of 3 cycles. This ensures that there will not be any resource conflict. It also contains eleven 16-bit registers for storing internal data.The datapath is controlled by a pipe-lined controller[7,8] which is depicted in Fig8 . The weights and the biases of all the neurons are stored in 4 ROMs (16x16 bit). The tansigmoidal activation function is implemented as a LUT ROM (64x16bit). Sixty four samples are taken in the range [0, 2.0] and the corresponding odd symmetric tansigmoid function values are calculated in Q15 format and stored in the ROM. A small logic is written to return maximum positive number in Q15 if the LUT input is greater than or equal to 2.0. For negative input, the LUT output is received after the 2's complement conversion.

The pipeline controller expects an active high interrupt from the data acquisition circuit whenever the pressure and temperature data are sampled. The controller in response initializes the datapath and the sequence logic in the controller as shown in Fig8 and it issues control signal vectors to the datapath in proper sequences for 12 consecutive segments of computation. The sequence logic implements the 3 cycle delay between successive P1 stages with the help of a 9-stage single bit shift register. It shifts in a repetitive sequence of "100" in to the shift register. Individual stages of shift registers qualify the corresponding control word. Finally the 9 qualified control words are OR'ed to generate the effective control word.

After the completion of P9 of the last computational segment, the controller generates an interrupt and enters an idle mode, waiting for next external interrupt. The total operation takes 9+11*3=42 pipeline cycles. The design assumes that the time interval between 2 consecutive interrupts is more than 42 cycles.

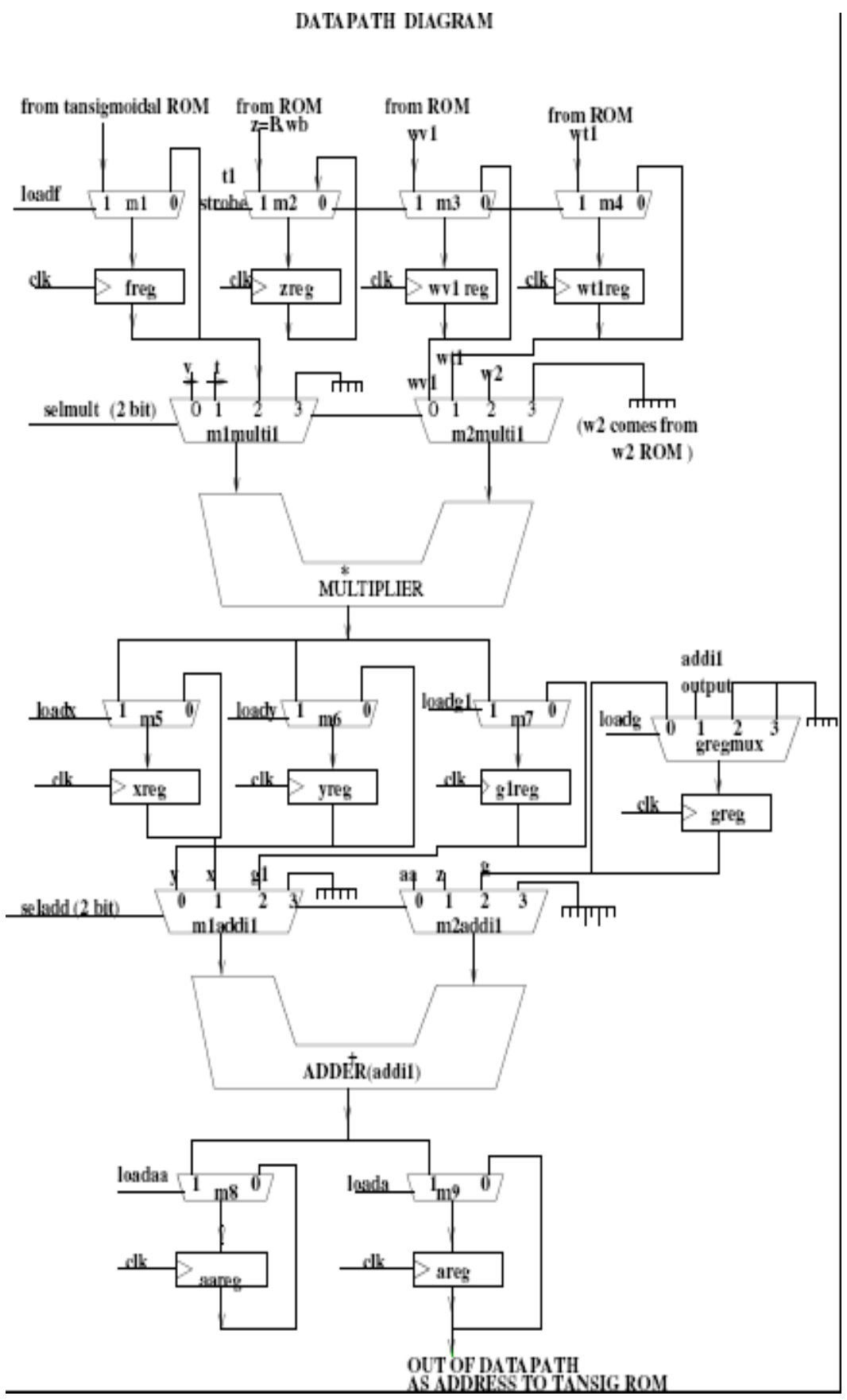

**Figure 7 : ANN datapath**

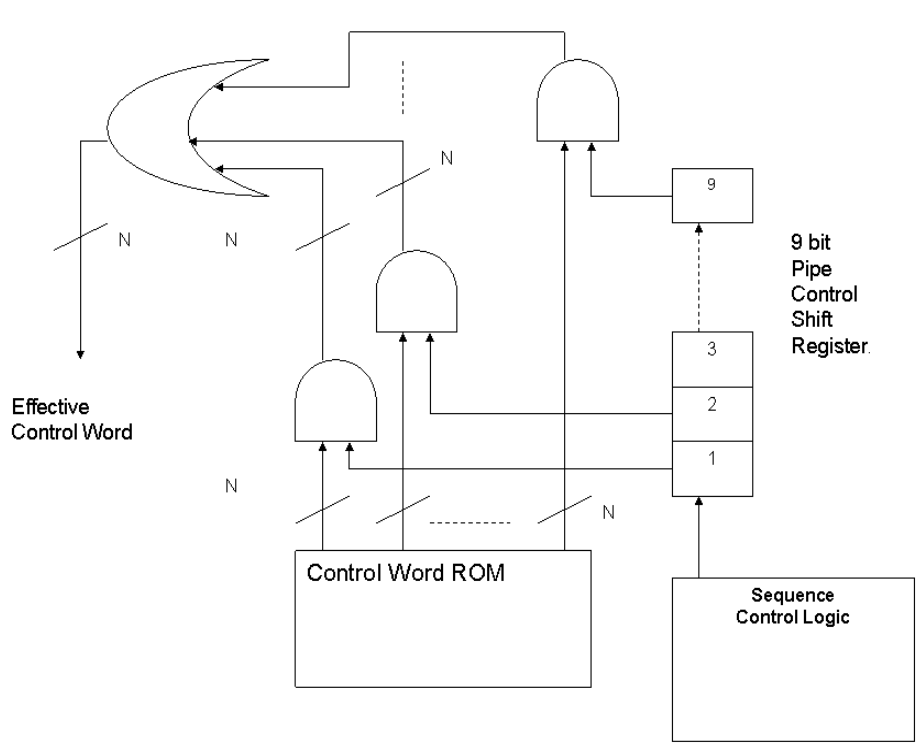

**Figure 8 : ANN Pipeline Controller**

The pipelined architecture is implemented on Xilinx (ISE 7.1) based FPGA platform and the device used is xc3s200 which is a Spartan chip. The functional simulation has been done using Modelsim simulator. The results of the trained network after FPGA implementation matches well with that of the MATLAB simulation studies. The normalized pressure and temperature values re acquired with the help of an ADC and the pressure value is displayed in a seven segment display. The synthesis results of the design are shown in Table 2 and timing results in Table 3.The normalized pressure and temperature values re acquired with the help of an ADC and the pressure value is displayed in a seven segment display. The synthesis results of the design are shown in Table 2 and timing results in Table 3 respectively. Table 2 gives the area overhead of the design indicating the usage of the CLBs of the Spartan 3 FPGA used in logic utilization, logic distribution and total number of look-up tables used for the design. It has also been observed that the pipelined based architecture based design enables the optimum utilization of CLB’s in the Spartan 3 chip which was not possible using simple behavioral modeling .

**TABLE 2: Synthesis Summary**

| **Logic Utilization** | **Used** | **Available** | **Utilization** | **Total number 4 input LUTs** | **Used** | **Available** | **Utilization** |
|---|---|---|---|---|---|---|---|
| Number of Slice Flip Flops: | 206 | 3,840 | 5% | Number used as logic | 441 | | |
| Number of 4 input LUT’s: | 441 | 3,840 | 11% | Number used as shift registers | 1 | | |
| **Logic Distribution :** | | | | No of bonded IOBs | 53 | 173 | 30 % |
| No. of occupied Slices | 322 | 1920 | 16% | No of Block RAMs | 5 | 12 | 41 % |
| No. of Slices containing only related logic: | 322 | 256 | 100 % | No. of MULT 18X18s | 1 | 12 | 8% |
| No. of Slices containing unrelated logic | 0 | 256 | 0% | No. of GCLKs | 1 | 8 | 12 % |

**TABLE 3: Timing results**

| Internal Path Group | | |
|---|---|---|
| **Budget** | **Clock Uncertainty** | **Worst Case Slack** |
| 15ns | 1ns | 1.523ns |
| **Input Path Group** | | |
| **Budget** | **Clock Uncertainty** | **Worst Case Slack** |
| 7ns | 1ns | 0.210ns |
| **Output Path Group** | | |
| **Budget** | **Clock Uncertainty** | **Worst Case Slack** |
| 10ns | 1ns | 2.467ns |

## 7. Conclusion

In this paper, FPGA based design and implementation of neural network compensators for nanocrystalline porous silicon MEMS pressure sensor has been reported in the pressure range 0-1 bar and temperature range 25$^o$C to 80$^o$C. It has been observed from the MATLAB simulation studies that the ANN based compensators reduces the error level by 98%. The neural network required 12 neurons in the hidden layer and the weights have been adjusted offline . The simulated structure of ANN has been implemented on a FPGA chip using pipelined architecture with a worst-case delay of 13.477ns.The datapath for the pipeline has been optimized for area so that the entire system can be efficiently realized on a FPGA chip. It has also been observed that the pipelined based architecture based design enables the optimum utilization of CLB's in the Spartan 3 chip which was not possible using simple behavioral modeling .